\documentclass[sigconf]{acmart}

\usepackage{algorithmic}
\usepackage{algorithm}
\usepackage{threeparttable}
\AtBeginDocument{%
  }

\begin{document}

\title{In-Memory ADC-Based Nonlinear Activation Quantization for Efficient In-Memory Computing}


\author{Shuai Dong$^{1}$,
        Junyi Yang$^{1}$, 
        Biyan Zhou$^{1}$,
        Hongyang Shang$^{1}$,
        Gourav Datta$^{2}$,
        Arindam Basu$^{1}$\\}
\affiliation{
 $^{1}$Department of Electrical Engineering, City University of Hong Kong, Hong Kong\
\country{China}\\
}
\affiliation{
$^{2}$ Department of Electrical, Computer, and Systems Engineering, Case Western Reserve University\
\country{USA}\\
}

\thanks{
This work was supported by the Research Grants Council of the HK SAR, China (Project No. CityU 11212823 and HKU C7003-24Y) and Innovation Technology Fund Mid-Stream Research Program (ITF MSRP) under Grant ITS/018/22MS. Any opinions, findings, conclusions, or recommendations expressed in this material do not reflect the views of the Government of the Hong Kong Special Administrative Region, the Innovation and Technology Commission, or the Innovation and Technology Fund Research Projects Assessment Panel. (Corresponding authors: Junyi Yang; Arindam Basu. e-mail: junyiyang8-c@my.cityu.edu.hk; arinbasu@cityu.edu.hk.)}








\renewcommand{\shortauthors}{Dong et al.}

\begin{abstract}
In deep networks, operations such as ReLU and hardware-driven clamping often cause activations to accumulate near the edges of the distribution, leading to biased clustering and suboptimal quantization in existing nonlinear (NL) quantization methods. This paper introduces Boundary Suppressed K-Means Quantization (BS-KMQ), a novel NL quantization approach designed to reduce the resolution requirements of analog-to-digital converters (ADCs) in in-memory computing (IMC) systems. By suppressing boundary outliers before clustering, BS-KMQ achieves more balanced and informative NL quantization levels. The resulting NL references are implemented using a reconfigurable in-memory NL-ADC, achieving a $7\times$ area improvement over prior NL-ADC designs. When evaluated on ResNet-18, VGG-16, Inception-V3, and DistilBERT, BS-KMQ achieves at least $3\times$ lower quantization error compared to linear, Lloyd–Max, cumulative distribution function (CDF), and K-means methods. It also improves post-training quantization accuracy by up to 66.8\%, 25.4\%, 66.6\%, and 67.7\%, respectively, compared to linear quantization. After low-bit fine-tuning, BS-KMQ maintains competitive accuracy with significantly fewer NL-ADC levels (3/3/4/4b). System-level simulations on ResNet-18 (6/2/3b) demonstrate up to a $4\times$ speedup and $24\times$ energy efficiency improvement over existing IMC accelerators.

\end{abstract}




\maketitle

\section{Introduction}
In-memory computing (IMC) has emerged as an effective way to alleviate the memory wall of von Neumann architectures. While this greatly reduces data movement and improves parallelism, maintaining high model accuracy typically requires moderate-to-high ADC resolution \cite{kim2024scaling}, leading the ADCs in IMC architectures to dominate system energy, area, and latency \cite{dong2025topkima}. Recent IMC accelerators therefore favor low-resolution (3–6 bit) ADCs combined with uniform quantization \cite{aguirre2024hardware}, which is hardware-friendly but mismatches the highly non-uniform distributions of neural activations. Consequently, linear quantization produces suboptimal decision levels and incurs severe accuracy loss at low bit widths. 

To better exploit activation statistics, various nonlinear (NL) quantization schemes—such as Lloyd–Max~\cite{cai2019low}, cumulative distribution function (CDF)~\cite{sun2020energy}, and K-means clustering~\cite{wu2025kllm}—adapt quantization boundaries to the activation distribution and can reduce quantization error compared to uniform quantization. However, Lloyd–Max quantization requires extensive iterative optimization and produces irregular and hardware-unfriendly step sizes~\cite{choi2016towards}; CDF-based approaches are highly sensitive to distribution outliers, often resulting in suboptimal quantization problem~\cite{zhai2021doro}; and standard K-means clustering suffers from boundary instability, especially near the distribution tails. This instability is particularly exacerbated in deep networks where the ReLU activation function accumulates a large number of outputs near zero (due to its non-negative nature) \cite{sun2020energy,wu2025kllm}, and hardware deployments typically impose value clamping to limit the activation range for compatibility with low-resolution ADCs. Moreover, the dynamically adjusted, data-driven boundaries produced by these methods remain difficult to implement efficiently in practical hardware systems.

On the hardware side, recent work has demonstrated NL quantization using specialized ADCs with nonvolatile devices. Yeo et al. designed a power-of-two SAR ADC with conversion skipping, achieving high energy efficiency by allocating finer resolution to dense input regions~\cite{yeo2024dynamic}, but its fixed quantization levels lack flexibility and cause measurable accuracy loss. Hong et al. proposed a memristor-based adaptive ADC using programmable CAM cells for data-driven NL quantization~\cite{hong2025memristor}, yet it suffers from scalability limits due to exponential Q-cell growth and sensitivity to device variation. Both designs share common NVM-related issues—device variability, limited endurance, and integration complexity—and are realized as peripheral ADC macros rather than the in-memory converters.
In parallel, in-memory ADC (IMA) designs have mainly used SRAM-based arrays with ramp ADCs for linear quantization only~\cite{yang2025high}. RRAM-based IMAs have been explored to embed NL activations into the ADC~\cite{yang2025efficient}, but they typically use the same activation profile across all layers and do not exploit data-dependent or layer-specific quantization.

Motivated by these limitations, this work presents Boundary Suppressed K-Means Quantization (BS-KMQ), an NL quantization method tailored for low-bit SRAM-based IMC. BS-KMQ explicitly suppresses boundary outliers induced by ReLU and clamping before clustering, yielding more informative quantization levels. The resulting NL  references are programmed into a reconfigurable in-memory (IM) NL-ADC. Our main contributions are summarized as follows:

\begin{figure}[t]
  \centering
  \includegraphics[width=0.9\linewidth]{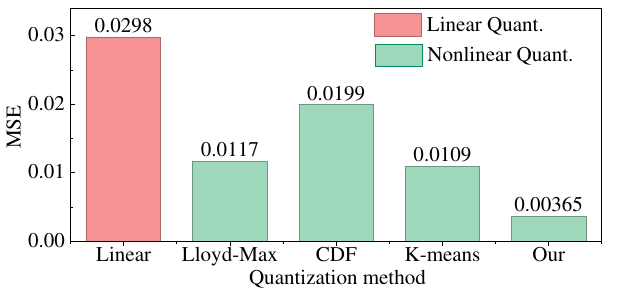}
  \caption{Mean Squared Error (MSE) comparison between linear \cite{yang2025high} and NL quantization schemes, including Lloyd--Max\cite{cai2019low}, CDF-based\cite{sun2020energy}, K-means\cite{wu2025kllm}, and our BS-KMQ method. All quantizers (3-bit) are evaluated on the activations from the first Conv-BN-ReLU block of ResNet-18 on Cifar-10 dataset.}
  \label{fig:motivation}
\end{figure}

\begin{itemize}
    \item BS-KMQ is proposed as a hardware-aware NL quantization scheme that suppresses boundary outliers and achieves $3\times$–$8\times$ lower quantization error than linear, Lloyd--Max, CDF-based, and standard K-means methods under 3-bit ADC precision (in Fig.~\ref{fig:motivation}).
    \item A reconfigurable (1-7 bits) IM NL-ADC architecture is designed to efficiently implement BS-KMQ references, enabling multi-bit NL quantization without complex analog circuitry. The overhead of NL-ADC (NL-ADC area/MAC array area) is merely 3.3\%, which is 7$\times$/5.2$\times$ improvement compared to 27\% for traditional NL ramp ADC in \cite{yang2025efficient} , and 17\% for linear SAR ADC in \cite{yin2025hybrid}. SPICE simulation results confirm the robustness of the IM NL-ADC architecture against process variations, increasing error only 1.2$\times$ under SS process corner relative to the TT corner, due to replica biasing.
    \item BS-KMQ is evaluated on CNNs (ResNet-18, VGG-16, Inception-V3) and a transformer (DistilBERT), improving up to 66.8\%/\allowbreak25.4\%/\allowbreak66.6\%/\allowbreak67.7\% higher post-training quantization (PTQ) accuracy than linear quantization. After low-bit finetuning, it maintains competitive accuracy with only 0.3\%/\allowbreak 0.5\%/\allowbreak 0.8\%/\allowbreak 1.2\% accuracy loss while using just 3/3/4/4-bit NLADC.
    \item System-level evaluation on Resnet-18 (6/2/3b) achieves up to 4$\times$ speedup and 24$\times$ energy-efficiency improvement compared to existing IMC accelerator.
\end{itemize}

\begin{algorithm}[!t]
\caption{Boundary Suppressed K-means Quantization (BS-KMQ)}
\label{alg:kmeans}
\begin{algorithmic}[1]
\STATE \textbf{Input:} Calibration batches $\{A^{(t)}\}_{t=1}^T$, bit-width $b$, tail ratio $\alpha = 0.005$
\STATE \textbf{Output:} Quantization centers $C=\{C_0, ...,C_{2^b-1}\}$

\STATE \vspace{2pt}\textbf{Stage 1: Robust statistical calibration}
\STATE Initialize sample buffer $S \gets \emptyset$
\FOR{$t = 1$ to $T$}
    \STATE Collect activations $A^{(t)}$
    \STATE Compute the $\alpha$ and $(1-\alpha)$ percentiles of $A^{(t)}$: $p_{\text{low}}, p_{\text{high}}$
    \STATE Keep central samples $A^{(t)}_{\text{cent}} = \{a \in A^{(t)} \mid p_{\text{low}} \le a \le p_{\text{high}}\}$
    \STATE $b_{\min} \gets \min(A^{(t)}_{\text{cent}})$,\quad $b_{\max} \gets \max(A^{(t)}_{\text{cent}})$
    \IF{$t == 1$}
        \STATE $g_{\min} \gets b_{\min}$,\quad $g_{\max} \gets b_{\max}$
    \ELSE
        \STATE $g_{\min} \gets 0.9 \cdot g_{\min} + 0.1 \cdot b_{\min}$
        \STATE $g_{\max} \gets 0.9 \cdot g_{\max} + 0.1 \cdot b_{\max}$
    \ENDIF
    \STATE Append $A^{(t)}_{\text{cent}}$ to $S$
\ENDFOR

\STATE \vspace{2pt}\textbf{Stage 2: Boundary-suppressed K-means clustering}
\STATE Clamp all samples in $S$ to $[g_{\min}, g_{\max}]$
\STATE Remove samples equal to $g_{\min}$ or $g_{\max}$ from $S$
\STATE Apply K-means to the remaining samples in $S$ to obtain $2^b - 2$ quantization center $C_q$
\STATE $C \gets \{g_{\min},C_q, g_{\max}\}$
\STATE \textbf{return} $C$
\end{algorithmic}
\end{algorithm}

\section{Methods}
\subsection{Nonlinear quantization: BS-KMQ}
\label{sec:bs-kmq}
To maintain accuracy in IMC systems under low-bit ADC resolution, we propose BS-KMQ, a hardware-aware NL quantization method that automatically adapts to the distribution of layer-wise activations. BS-KMQ explicitly suppresses boundary outliers induced by ReLU and clamping before clustering, so that the limited quantization levels are allocated to the informative interior region of the distribution, leading to significantly lower quantization error under low-bit quantization.

BS-KMQ proceeds in two stages: (i) robust statistical calibration and (ii) boundary-suppressed K-means clustering for reference generation. During calibration, we feed the calibration dataset through the network in mini-batches. For each batch, we first discard the top and bottom 0.5\% of activation values, i.e., we retain only the central 99\% of samples and treat the extreme 0.5\% tails on both sides as outliers. From the remaining samples of each batch, we compute a batch-wise minimum and maximum ($b_{\min}$, $b_{\max}$), and update the global boundary range $(g_{\min}, g_{\max})$ using an exponential moving average (EMA) \cite{klinker2011exponential}:
\begin{equation}
\begin{aligned}
    g_{\min} &= 0.9 \cdot g_{\min} + 0.1 \cdot b_{\min}, \\
    g_{\max} &= 0.9 \cdot g_{\max} + 0.1 \cdot b_{\max}.
\end{aligned}
\end{equation}

\begin{figure*}[!t]
  \centering
  \includegraphics[width=0.9\linewidth]{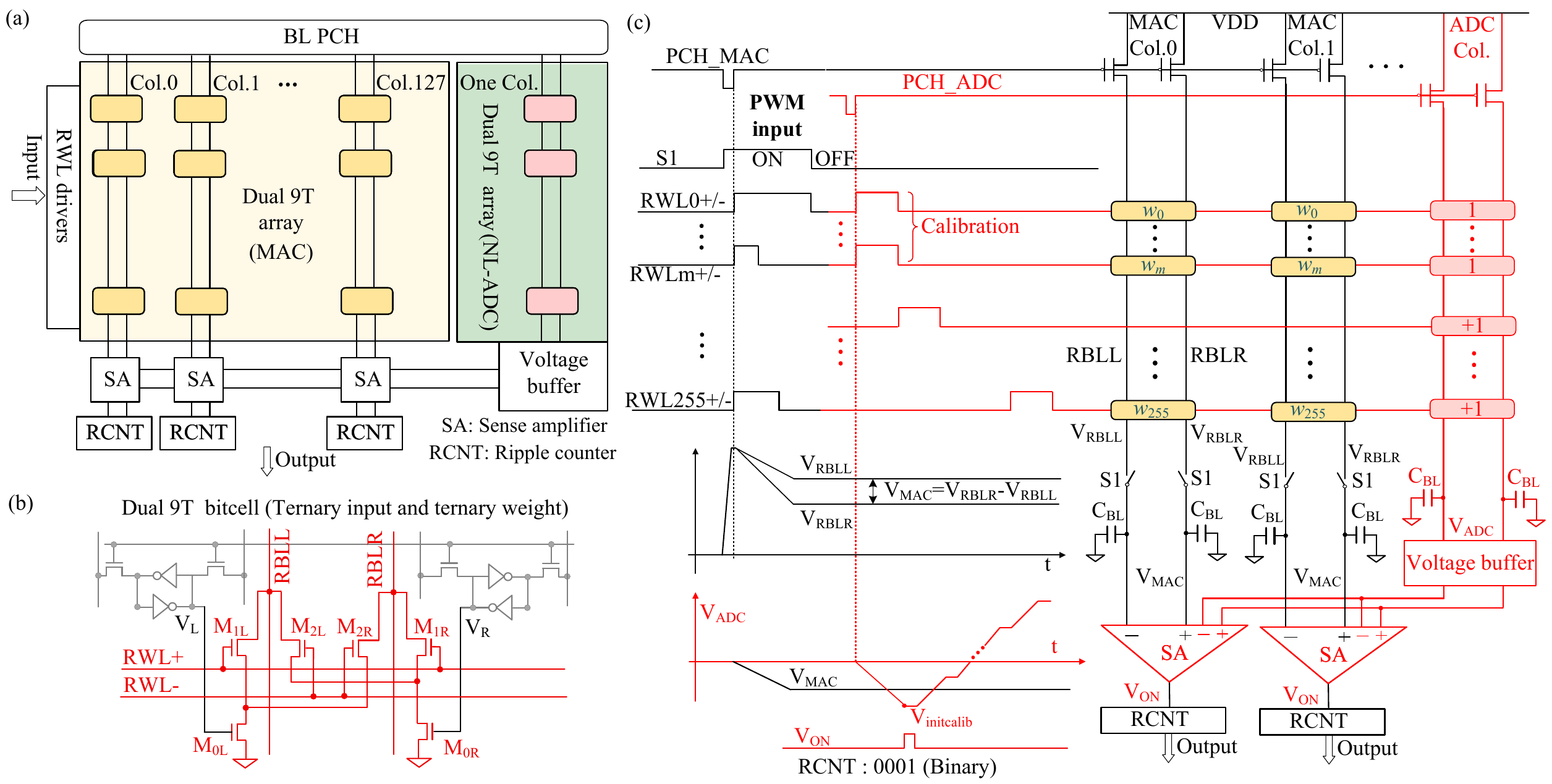}
  \caption{Overall structure of the proposed dual 9T SRAM IMC and IM NL-ADC: (a) System hardware structure. (b) Information of dual 9T SRAM bitcell. (c) Detailed circuit and timing diagram.}
  \label{fig：chip_overall_architecture}
\end{figure*}

This batch-wise EMA yields a robust layer-wise range that is insensitive to the outliers. After calibration, all activation samples used for clustering are clamped into $[g_{\min}, g_{\max}]$, and those saturating at the lower or upper bounds are treated as boundary outliers. These boundary-clamped samples are removed from the clustering pool so that K-means operates only on interior samples, avoiding biased centroids near the edges. Finally, we include $(g_{\min}, g_{\max})$ in the reference set to ensure full‑range coverage and compatibility with IM NL‑ADC implementations. The BS-KMQ procedure is summarized in Algorithm~\ref{alg:kmeans}.

We then obtain a set of quantized centers $C = \{C_0, C_1, \dots, C_{2^b-1}\}$. In the algorithmic view, each activation is quantized to the nearest center. However, in hardware, the ADC compares the input only against a set of reference levels and effectively implements a ``floor'' operation rather than an explicit nearest-center rounding. 

To emulate nearest-center quantization with such hardware, BS-KMQ converts the learned centers into a set of quantization reference levels $R$ according to
\begin{equation}
\begin{aligned}
    R_0 &= C_0, \quad i = 0 \\  
    R_i &= \frac{C_{i-1} + C_i}{2}, \quad i = 1,\dots,2^b - 1,
\end{aligned}
\label{eq:map}
\end{equation}
Each reference $R_i$ defines the lower bound of the quantization region corresponding to center $C_i$. When doing quantization, the ADC compares the input against $R_i$ and outputs the index $i$ of the largest reference level not exceeding the input. Then the index is mapped to the corresponding quantized value $C_i$, effectively realizing nearest-center quantization.
For example, in a 3-bit BS-KMQ configuration with centers
\[
C = \{0, 0.125, 0.25, 0.5, 1.0, 2.0, 4.0, 8.0\},
\]
the derived references become
\[
R = \{0, 0.0625, 0.1875, 0.375, 0.75, 1.5, 3.0, 6.0\}.
\]
An input of 0.05 falls below $R_1$ and maps to $C_0 = 0$, while an input of 0.07 lies between $R_1$ and $R_2$ and maps to $C_1 = 0.125$. This mechanism preserves the benefits of NL clustering while ensuring efficient implementation with floor-type ADCs. 

\subsection{Hardware architecture}
To implement the NL quantization, Fig. \ref{fig：chip_overall_architecture}(a) depicts an SRAM-based macro architecture employing IMC for MAC and ADC operations. This design integrates several key components: a $256\times 128$ computational crossbar composed of dual 9T SRAM bitcells for MAC operations, a $256\times 1$ bitcell array functioning as  shared reference bitcells for a reconfigurable (1-7 bit) IM NL-ADC with programmable step size to realize NL quantization, alongside peripheral circuits including read word line (RWL) drivers, sense amplifiers (SAs), voltage buffers, and ripple counters (RCNTs). 

The macro architecture is built upon the dual 9T SRAM bitcell, illustrated in Fig. \ref{fig：chip_overall_architecture}(b). This bitcell features a decoupled read path with six NMOS transistors (highlighted in red) to perform ternary multiplication between ternary input and ternary weight. Input polarity is determined by activating either the positive read word line (RWL+) for positive inputs or the negative read word line (RWL-) for negative inputs. The ternary weights are stored in the standard 6T-SRAM cells, encoded in three states: -1 ($V_{L}$=L, $V_{R}$=H); 0 ($V_{L}$=L, $V_{R}$=L); and +1 ($V_{L}$=H, $V_{R}$=L). When a weight of 0 is stored, neither the left nor right read bit lines (RBLL or RBLR) discharge, regardless of the input. Zero-valued weights create no discharge paths within the bitcell, thereby reducing the energy consumed by RBL discharge. For non-zero weights (-1 or +1), the corresponding read bit line discharges through the activated NMOS transistor gated by the asserted RWL. The result of the ternary multiplication is represented by the voltage difference between the right and left RBLs ($V_{MAC}=V_{RBLR}-V_{RBLL}$), as shown in Fig. \ref{fig：chip_overall_architecture}(b). The compact layout of this dual 9T bitcell, designed in a 65 nm process, occupies an area of 3.6 $\mu m$$\times$1.9 $\mu m$.

Fig. \ref{fig：chip_overall_architecture}(c) illustrates the timing diagrams for both MAC operations and NL-ADC operations. During the computation phase, multi-bit inputs encoded as pulse-width modulation (PWM) signals are applied for the MAC operation when switch S1 (Fig. \ref{fig：chip_overall_architecture}(c) ) is turned ON. The MAC operation creates $V_{MAC}=V_{RBLR}-V_{RBLL}=\sum_{k=1}^N W_k X_k$ on the bitlines (using current-mode operation) where $N=256$ is the dimension of the input vector. 
Following the computation phase, switch S1 is turned OFF, allowing the bitline capacitors ($C_{BL}s$) to maintain the $V_{MAC}s$  representing the MAC results. 
Simultaneously, a reference‑generation circuit within the NL‑ADC column produces a ramp voltage ($V_{ADC}$), whose step size is designed as $R_{i+1}-R_i$. This $V_{ADC}$ is shared by all 128 SAs to perform concurrent comparisons between the $V_{ADC}$ and all $V_{MAC}s$. We employ the voltage buffer and double differential SA from \cite{yang2025high} to implement the NL-ADC with shared references. The key difference is that our ADC employs a NL architecture, whereas the ADC in \cite{yang2025high} is linear. This results in the requirement for different numbers of bitcells when implementing ADCs of the same resolution. We analyze this overhead in the following section. The resulting thermometer-coded outputs from SAs are subsequently converted to binary format through RCNTs. Detailed descriptions of NL-ADC circuit block are provided in the following subsection.

\subsection{Reconfigurable  In-memory Nonlinear ADC architecture}
  \label{section：NLADC}
\begin{figure}[t]
  \centering
  \includegraphics[width=0.9\linewidth]{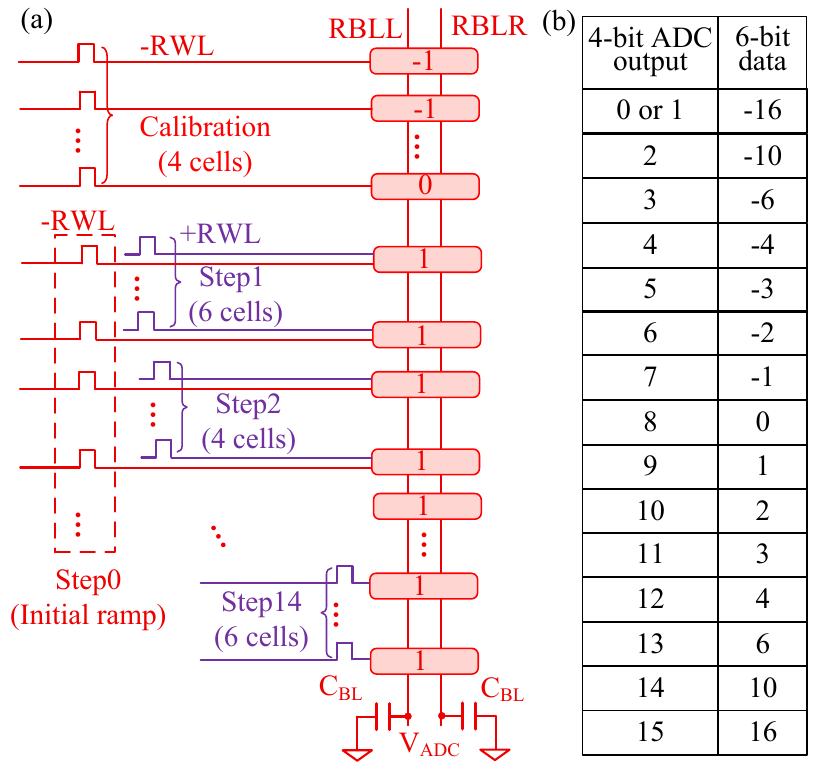}
  \caption{(a) Schematic and timing of generating references of 4-bit NL-ADC. (b) The mapping relationship from the 4-bit output of the NL-ADC to the actual 6-bit data.}
  \label{fig：NLADC}
\end{figure}

The architecture and operational principle of the IM NL-ADC are shown in Fig. \ref{fig：chip_overall_architecture}(c) (red part). Notably, the replica bitcells utilized for ADC operation are identical to those in the MAC section. The operational sequence is conceptually divided into two distinct phases: (1) Generation of an initial ramp voltage $V_{\text{initcalib}}$ (Fig. \ref{fig：chip_overall_architecture}(c), bottom). In this configuration, many -RWL signals are activated simultaneously, causing negative $V_{\text{initcalib}}$; and (2) Generation of the subsequent $V_{\text{ADC}}$, achieved by simultaneously enabling many bit-cells with  weights of $+1$ in every step as shown in Fig. \ref{fig：NLADC}(a). Fig. \ref{fig：chip_overall_architecture}(c) (bottom) schematically illustrates the generation of the global reference voltage ($V_{\text{ADC}}$) for NL-ADC.   It is worth noting that the previous IM ADCs (\cite{yang2025high},\cite{kim65nm8TSRAMIMC},\cite{KIMTCASIvoltagemode}) are constrained by its binary input format (only one RWL in every column). To generate the initial ramp voltage, it must employ a separate array of many bitcells. Consequently, this method suffers from a large area overhead, with the the initial ramp voltage generation circuit consuming an area comparable to the ADC core itself (IM ADC area/initial ramp area $\approx$50\%), making the cost prohibitively high. In contrast to conventional implementations, the proposed architecture eliminates these bitcells through the dual 9T bitcell design. This bitcell supports both positive (RWL+) and negative (RWL-) inputs, where RWL- serve to generate the initial ramp voltage while RWL+ as shown in Fig. \ref{fig：NLADC}(a) produce the increasing $V_{\text{ADC}}$. 

\begin{figure}[t]
  \centering
  \includegraphics[width=\linewidth]{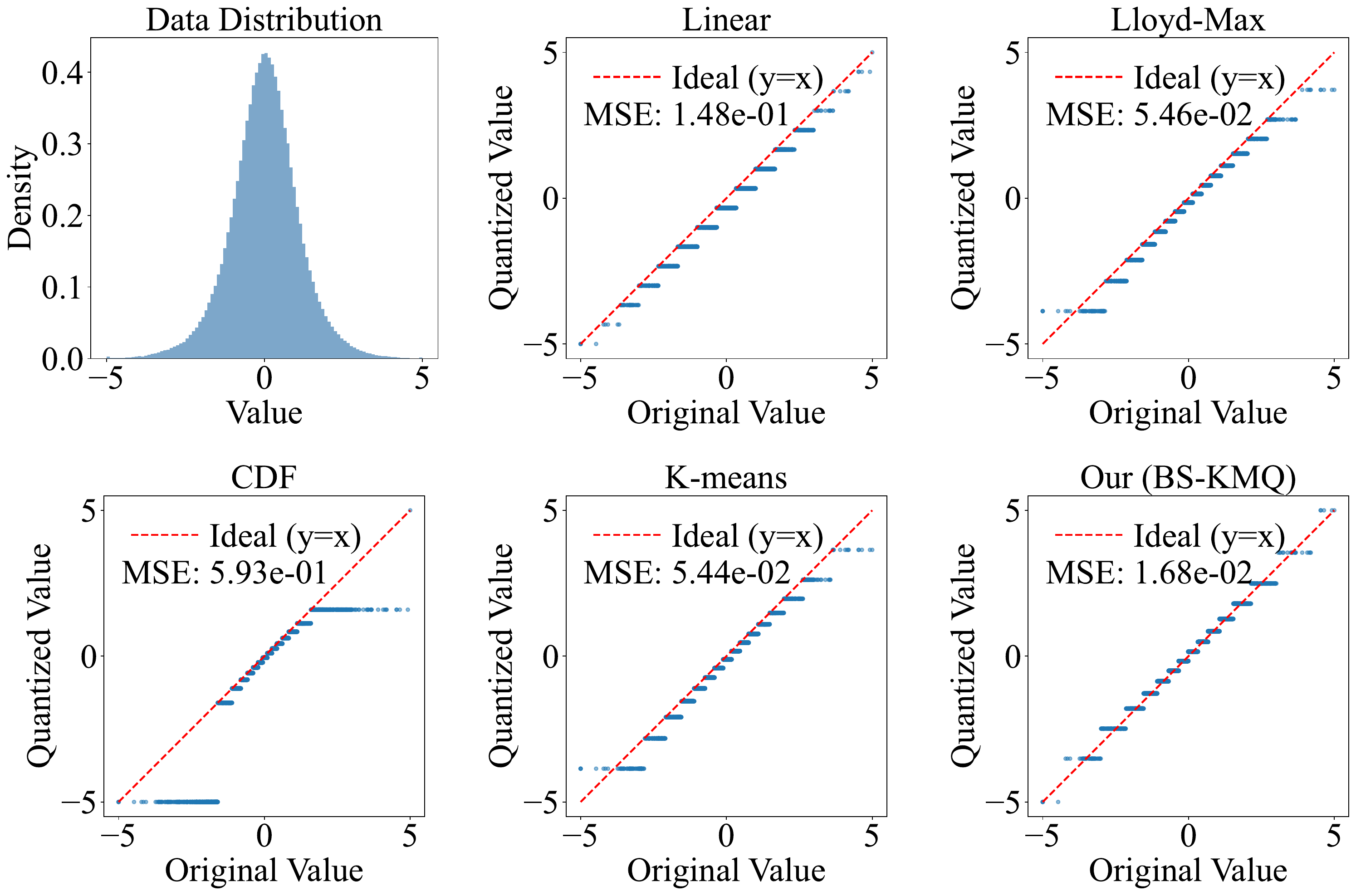}
  \caption{Analysis of MSE for 4-Bit Quantization in a representative layer of DistilBERT. Our method is compared with conventional linear quantization and several NL schemes, including Lloyd–Max, CDF, and standard K-means.}
  \label{fig:transformer MSE}
\end{figure}

 We can generate reference voltages with different step sizes of NL-ADC by enabling varying numbers of bitcells as shown in Fig. \ref{fig：NLADC}(a), thereby achieving flexible quantization of MAC values with different data distributions. The actual generated $V_{\text{ADC}}$ deviates from its ideal characteristic due to hardware non-idealities. To mitigate this, a zero-crossing calibration technique  \cite{yang2025efficient}  is employed to adjust $V_{\text{initcalib}}$, ensuring $V_{\text{ADC}}$ crosses the zero point and thereby reducing error. The calibration circuitry requires four additional bitcells, as shown in Fig. \ref{fig：NLADC}(a), leaving 252 bitcells (256-4) available for NL-ADC. For the 4-bit NL-ADC implementation depicted in Fig. \ref{fig：NLADC}(a), only 32 bitcells are required (excluding the four calibration bitcells). Consequently, our NL-ADC can be flexibly configured for higher resolutions, with a maximum achievable resolution of 7 bits.  Compared to linear IM ADC in \cite{yang2025efficient}, our IM NL-ADC requires more bitcells at the same resolution. For example, we need 32 bitcells while a linear IM ADC only requires 16 bitcells for a 4-bit output.

To minimize the MSE in NL quantization, as discussed in Sec.~\ref{sec:bs-kmq}, we establish a mapping between the 4-bit NL-ADC output levels (the index of $R$) and the corresponding 6-bit quantized centers ($C$), based on the practical activation distributions (from first Conv-BN-ReLU block) observed by BS-KMQ. This mapping is illustrated in Fig.~\ref{fig：NLADC}(b).

\begin{figure}[!t]
  \centering
  \includegraphics[width=\linewidth]{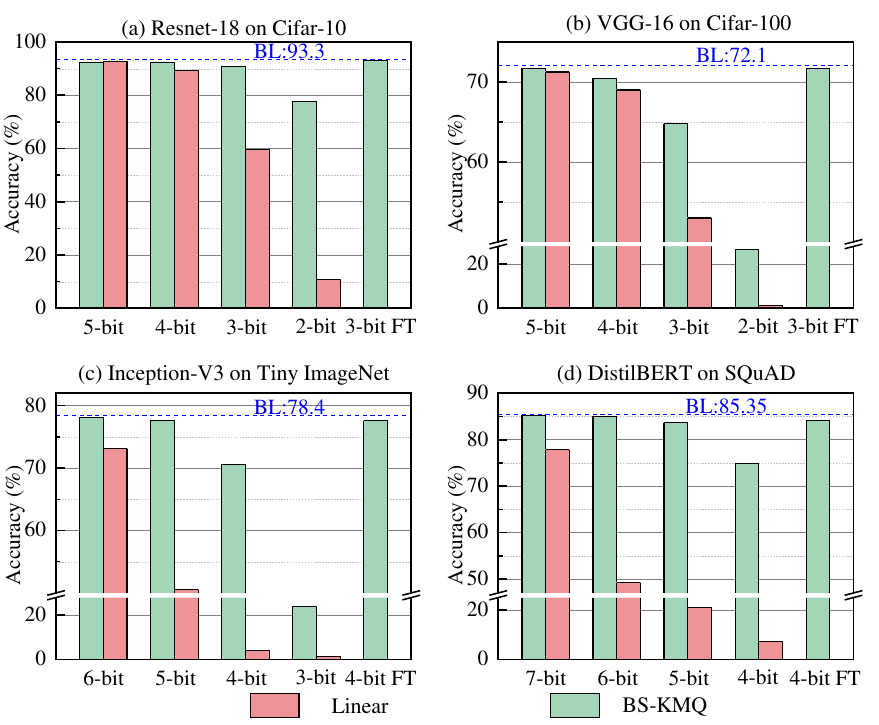}
  \caption{PTQ accuracy comparison between linear quantization \cite{yang2025high} and our method, along with fine-tuning (FT) accuracy, for (a) ResNet-18, (b) VGG-16, (c) Inception-V3, and (d) DistilBERT. BL denotes the floating-model baseline.}
  \label{fig:post_acc}
\end{figure}

\section{Simulated Results}
\subsection{Software performance}
\textbf{MSE of BS-KMQ on CNN and transformer:} To evaluate the effectiveness of the proposed BS-KMQ method, we first compare it against a conventional linear quantization \cite{yang2025high} and several representative NL schemes, including Lloyd-Max\cite{cai2019low}, CDF\cite{sun2020energy}, and standard K-means\cite{wu2025kllm}. We analyze both convolutional and transformer models: specifically, the activations after the first Conv-BN-ReLU block in ResNet-18 on CIFAR-10 and the query projection $Q = W X$ in the first attention layer of DistilBERT on SQuAD. For each layer, we apply the above quantization methods and measure MSE between the original and quantized activations.

As shown in Fig.~\ref{fig:motivation} and Fig.~\ref{fig:transformer MSE}, BS-KMQ consistently achieves the lowest quantization error among all baselines. On ResNet-18, it reduces the MSE by approximately $3\times$–$8\times$ compared to other methods, while on DistilBERT it yields up to $3\times$–$35\times$ lower MSE. This improvement mainly comes from explicitly suppressing boundary outliers and learning more informative quantization references, which are then mapped to the appropriate quantized center, enabling a closer approximation to the true activation distribution under low-bit ADC constraints.

\textbf{PTQ and Fine-tuning (FT) accuracy:} We then compare the PTQ accuracy of linear quantization and BS-KMQ on four representative benchmarks: ResNet-18 on CIFAR-10, VGG-16 on CIFAR-100, Inception-V3 on Tiny-ImageNet, and DistilBERT on SQuAD. As shown in Fig.~\ref{fig:post_acc}, BS-KMQ improves accuracy by up to 66.8\%, 25.4\%, 66.6\%, and 67.7\% over linear quantization on these four tasks at the same bit width. After low-bit finetuning, BS-KMQ maintains competitive accuracy while using only 3/3/4/4-bit ADC levels, with absolute accuracy drops of just 0.3\%/0.5\%/0.8\%/1.2\% for the four models, respectively.

\begin{figure}[t]
  \centering
  \includegraphics[width=\linewidth]{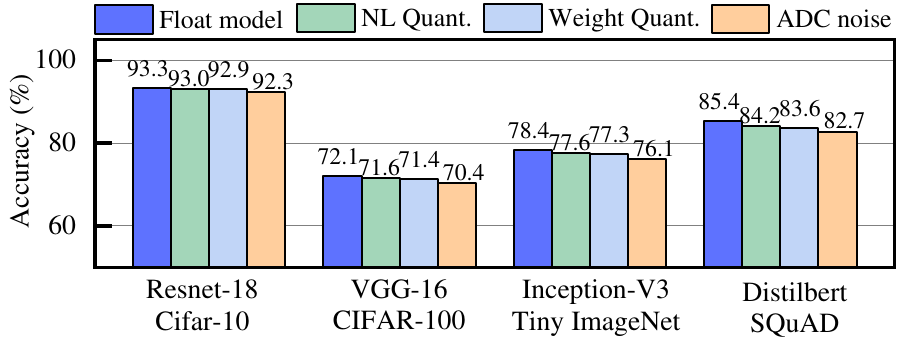}
  \caption{Quantization effects and ADC impact on model accuracy for ResNet-18, VGG-16, Inception-V3, and DistilBERT.}
  \label{fig:full_acc}
\end{figure}

\textbf{Weight quantization:} To implement these models in hardware, weight quantization is also required. Since this work focuses on NL quantization of activations and the weight ranges are fixed after training (unlike dynamically varying activations), weights are easier to handle and we simply adopt linear quantization. The weights of the four models are quantized to 2/3/4/4b, respectively, incurring only 0.10\%/0.22\%/0.37\%/0.51\% accuracy loss, as shown in Fig.~\ref{fig:full_acc}. 

\begin{figure}[t]
  \centering
  \includegraphics[width=1.02\linewidth]{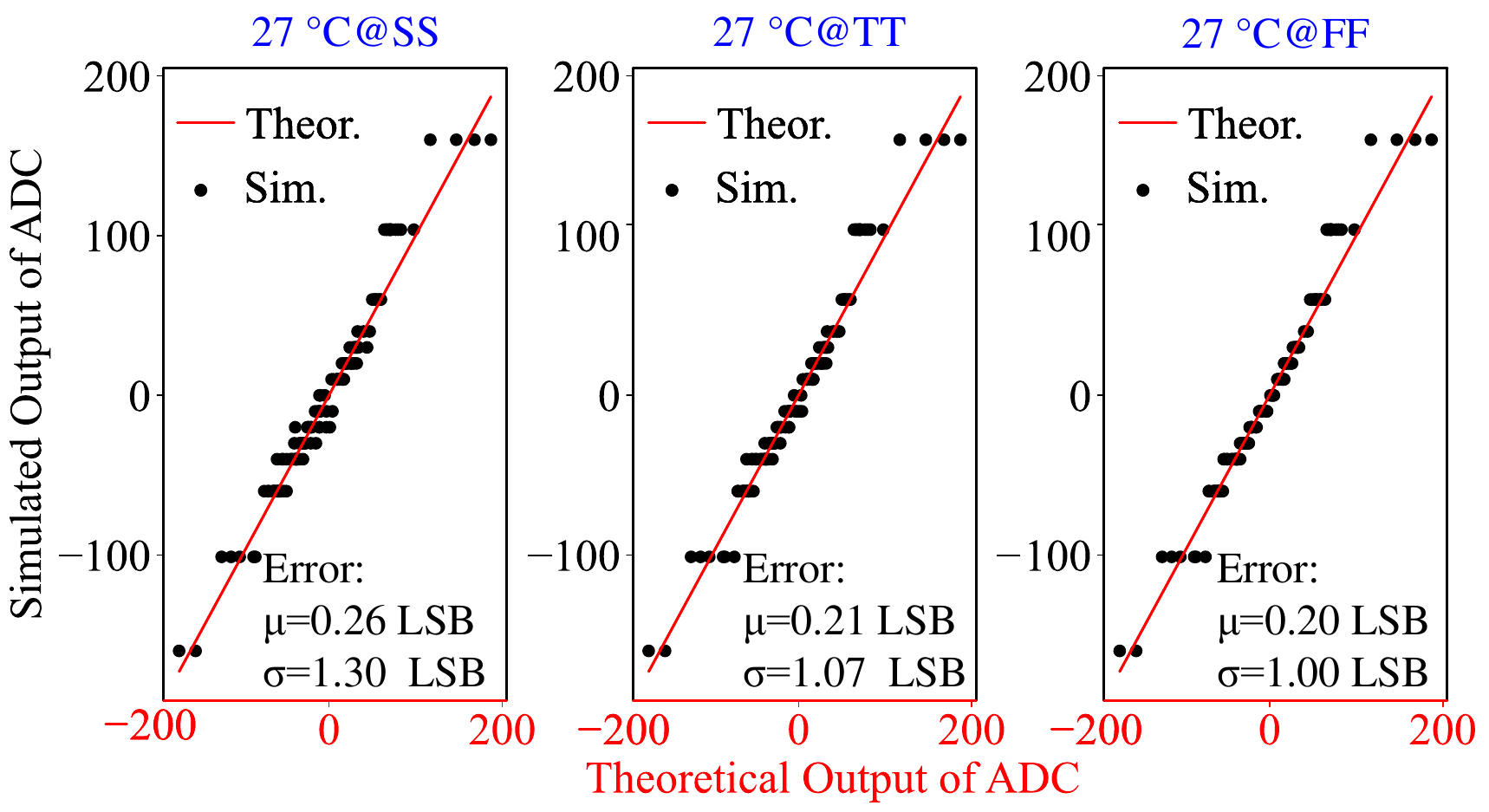}
  \caption{Simulated results with 6-bit input and 4-bit output: NL-ADC output vs.theoretical result of MAC at different process corners (SS, TT, FF) (The minimum step size of NL-ADC is 10).}
  \label{fig:adc_noise}
\end{figure}

\textbf{ADC noise impact:} SPICE simulations are performed in a 65 nm CMOS process to evaluate the proposed IM NL-ADC architecture. The design employs two clock domains: a 200~MHz clock for PWM-based inputs and a 200~MHz clock for IMA operations. The 6T-SRAM array operates with a 1~V pre-charge voltage and a 0.45~V supply voltage. The simulated 4-bit ADC outputs with 2-bit weights are compared against ideal ADC outputs across different process corners (TT, FF, SS), as illustrated in Fig.~\ref{fig:adc_noise}. The resulting errors follow an approximately Gaussian distribution with consistently small mean ($\mu$) and standard deviation ($\sigma$), demonstrating that the IM NL-ADC is robust to process variations. The standard deviation of the IM NL-ADC error exhibits minimal degradation, showing merely a 1.2$\times$ increase under SS process corner relative to the TT corner. These results confirm the design's exceptional robustness against process variations, which is attributed to the implemented replica biasing technique.  The nominal error distribution at the TT corner, $\mathcal{N}(0.21, 1.07)$, is then used to inject ADC noise into the four quantized networks. As shown in Fig.~\ref{fig:full_acc}, the resulting accuracy degradation is limited to 0.6\%, 1.0\%, 1.2\%, and 0.9\% for ResNet-18, VGG-16, Inception-V3, and DistilBERT, respectively. Inception-V3 exhibits a slightly larger accuracy loss due to its deeper architecture, which causes noise to accumulate over more layers.

\begin{figure}[t]
  \centering
  \includegraphics[width=1.0\linewidth]{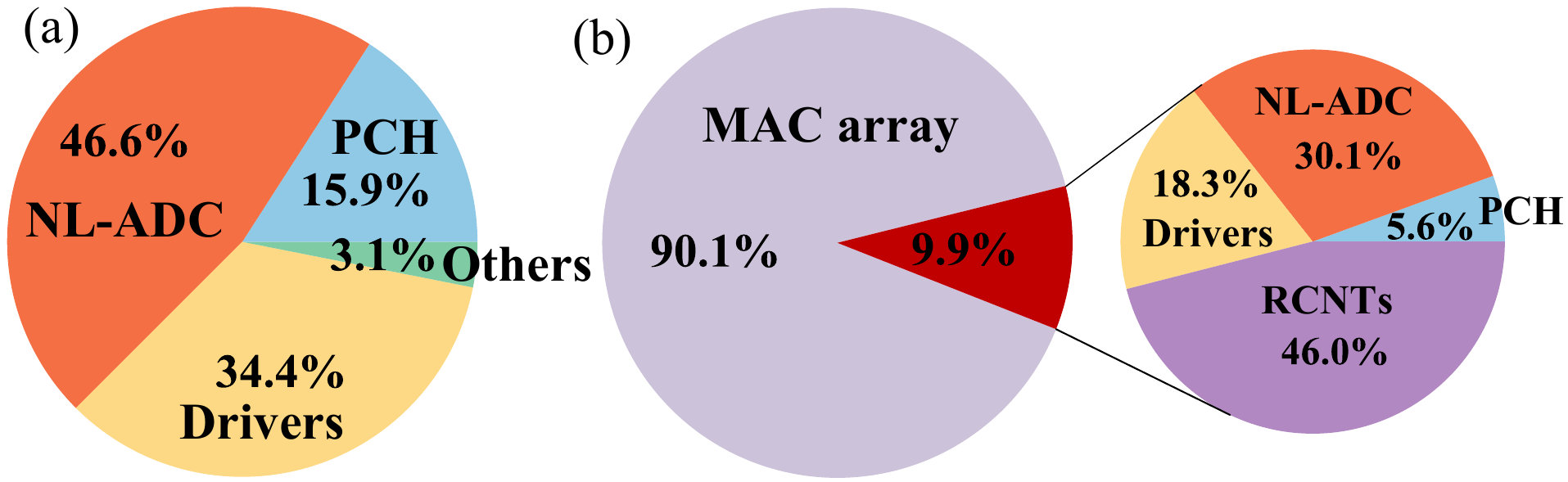}
  \caption{(a) Energy breakdown of macro (6/4-bit input/output and 2-bit weight). (b) Area breakdown of macro.}
  \label{energy_area}  
\end{figure}

\subsection{Hardware performance}
\textbf{Macro level evaluation:}
Fig. \ref{energy_area}(a) presents the simulated energy distribution of our design for a configuration with 4-bit/6-bit input/output and 2-bit weights. The analysis reveals that the NL-ADC and drivers operations dominate the total energy consumption. The area breakdown  of the macro  is shown in Fig. \ref{energy_area}(b), which occupies a total area of 0.248 mm\textsuperscript{2}. Notably, the 128 IM NL-ADCs exhibit a compact footprint, consuming only 3.3\% of the MAC array's area. The proposed design achieves a remarkable  7$\times$ improvement in area overhead (NL-ADC area/MAC array area) compared to the NL ramp ADC in \cite{yang2025efficient} (23\%).  The macro achieves 246 TOPS/W and 0.55 TOPS/mm\textsuperscript{2} at 6-bit input, 2-bit weight and 4-bit output.  Compared to the linear IM ADC \cite{yang2025efficient} with the same condition, our IM NL-ADC results in a $\approx$ 30\% increase in the energy consumption due to increasing bitcells as mentioned in Section \ref{section：NLADC} . 

To accommodate the computational demands of sophisticated neural networks, the proposed hardware architecture natively incorporates multi-bit weight representations. This capability is achieved through a parallel bitcell connection scheme, where the numerical value of a weight is represented by the number of concurrently activated bitcells. For example, a 4-bit weight implementation (excluding sign bit) utilizes a binary encoding approach where the three magnitude bits are mapped to parallel connections of 1, 2, and 4 identical bitcell structures (resulting in a total of 7 cells per 4-bit weight). The sign representation is inherently implemented through a symmetrical dual 9T cell architecture that differentially utilizes left and right computational paths. A key distinguishing feature of the proposed macro architecture is its high reconfigurability, supporting dynamic precision scaling across all data types: inputs (1-7 bits), weights (2-4 bits), and outputs (1-7 bits).

\textbf{System-level evaluation:} To account for system-level overheads, we perform end-to-end simulations of a ResNet-18 IMC accelerator on CIFAR-10. The costs of the crossbar and IM NL-ADC are obtained from SPICE simulations, while peripheral costs (including interconnect, buffers, and accumulation units, etc.) are estimated using NeuroSim \cite{peng2020dnn+}. For consistency with the SPICE simulations, the technology node of NeuroSim is also set as 65~nm technology node.  As summarized in Table~I, the proposed BS-KMQ–based accelerator achieves a throughput of 2 TOPS and an energy efficiency of 31.5 TOPS/W. Compared with three existing IMC systems—SRAM-based with linear ADC, RRAM-based with NL-ADC, and ferroelectric capacitive (FCA)–based with NL-ADC—our design achieves up to 4$\times$ speedup and 24$\times$ improvement in energy efficiency.

\begin{table}[t]
\centering
\caption{Comparison with state-of-the-art IMC designs}
\label{tab:comparison}
\begin{threeparttable}
\renewcommand{\arraystretch}{1.2}
\begin{tabular}{|c|c|c|c|c|}
\hline
\textbf{Metrics} & \begin{tabular}[c]{@{}c@{}}TCASI'24 \\ \cite{mao202428} \end{tabular} 
                 & \begin{tabular}[c]{@{}c@{}}VLSI’23\\ \cite{wen202328nm} \end{tabular}
                 & \begin{tabular}[c]{@{}c@{}}SSCL'24\\ \cite{yeo2024dynamic} \end{tabular} 
                 & \textbf{Ours} \\ \hline
Tech. (nm)              & 28        & 28           & 180              & 65     \\ \hline
Supp. (V)               & 0.9–0.95  & 0.7-0.8            & 1.8              & 1.1    \\ \hline
Freq. (MHz)             & 160–340   & 50-200             & 12               & 200    \\ \hline
Bitcell                 & 9T1C      & RRAM         & FCA              & Dual 9T \\ \hline
Array Size              & 1152×128  & -            & 16×8             & 256×128 \\ \hline
I/O bits                & 2,4,8 / 2,4,8 & 8/4       & 1.5/5            & 6/3  \\ \hline
W. bits                 & 2,4,8     & 8          & 1.5              & 2      \\ \hline
ADC Type                & Linear    & NL           & NL               & IM NL   \\ \hline
Reconfig.               & N         & N            & N                & Y   \\ \hline
Network                 & ResNet-18 & Resnet-20        & ResNet-18        & ResNet-18 \\ \hline
Dataset                 & CIFAR-10  & CIFAR-100     & CIFAR-10         & CIFAR-10 \\ \hline
Acc. loss (\%)\tnote{a} & 3.22      & 0.45         & 1.7              & 1.0   \\ \hline
TOPS                    & 0.52         & 0.34            & -                & 2.0     \\ \hline
TOPS/W\tnote{b}         & 5.45–21.82& 0.52-1.29            & 13.27-34.6        & 31.5     \\ \hline
\end{tabular}
\vspace{3pt}
\begin{tablenotes}
\footnotesize
\item[a ] Compared to software simulation of floating-point configuration.
\item[b] TOPS/W = Reported $\times$ (Tech./65nm) $\times$ (Supp./1.1V)$^2$.
\end{tablenotes}
\end{threeparttable}
\end{table}

\section{Conclusion}

This work presented BS-KMQ, an NL quantization strategy tailored for reducing ADC resolution in IMC system. By suppressing value accumulation near distribution boundaries—typically caused by ReLU and clamping—BS-KMQ enables more balanced and effective clustering. This approach significantly improves quantization quality, achieving at least $3\times$ lower error compared to linear, Lloyd–Max, and CDF and K-means methods. 
Extensive evaluations across ResNet-18, VGG-16, Inception-V3, and DistilBERT demonstrate that BS-KMQ improves up to 66.8\%, 25.4\%, 66.6\%, and 67.7\% higher PTQ accuracy over linear baselines. After low-bit FT, the proposed method maintains competitive accuracy with significantly fewer NL-ADC levels (3/3/4/4b).
At the system level, simulation on ResNet-18 (6/2/3b) achieves up to 4$\times$ speedup and $24\times$ energy efficiency improvement over prior IMC accelerators. These results highlight the promise of BS-KMQ as an effective and hardware-friendly quantization solution for energy-efficient neural network inference.

\newpage
\bibliographystyle{ACM-Reference-Format}
\bibliography{sample-base}

\end{document}